\documentclass[fleqn,twoside]{article}
\usepackage{espcrc2}

\usepackage{amssymb}
\usepackage{amsbsy}
\usepackage{amsfonts}

\usepackage{graphicx}
\newcommand{\AmS}{{\protect\the\textfont2
   A\kern-.1667em\lower.5ex\hbox{M}\kern-.125emS}}

\newcommand{\be}{\begin{equation}}
\newcommand{\ee}{\end{equation}}
\newcommand{\bea}{\begin{eqnarray}}
\newcommand{\eea}{\end{eqnarray}}

\title{
\vspace{-9mm}
\rightline{\small ITEP-LAT/2002-14, KANAZAWA-02-22, RCNP-Th02015}
A fresh look on the flux tube in Abelian-projected
SU(2) gluodynamics}

\author{
Y. Koma,\address{Institute for Theoretical Physics,
Kanazawa University, kakuma-machi, Kanazawa 920-1192,
Japan\\[-0.5em]}\thanks{Talk presented by Y. Koma
at Lattice 2002 symposium}
M. Koma,\address{RCNP,
Osaka University,  Mihogaoka 10-1, Ibaraki,
Osaka 567-0047, Japan\\[-0.5em]}
T. Suzuki,$^{\rm a}$
E.-M. Ilgenfritz,$^{\rm b}$
and M.I. Polikarpov\address{ITEP, B.Cheremushkinskaya 25,
RU-117259 Moscow, Russia}\\}

\begin{document}

\begin{abstract}

We reconsider the properties of the $Q\bar{Q}$ flux tube
within Abelian-projected SU(2) lattice gauge theory
in terms of electric field and monopole current.
In the maximal Abelian gauge we assess the influence
of the Gribov copies on the apparent flux-tube profile.
For the optimal gauge fixing
we study the independence of the profile on the lattice spacing for
$\beta=$ 2.3, 2.4, and 2.5115 on a $32^4$ lattice.
We decompose the Abelian Wilson loop
into monopole and photon parts and compare the electric
and monopole profile emerging from different
sources with the field strength and monopole current
within the dual Ginzburg-Landau theory.
\end{abstract}

\maketitle

The flux-tube profile in SU(2) lattice gauge theory in the
maximal Abelian gauge (MAG) has been studied in order to get
microscopic information on the quark confinement
mechanism~\cite{Singh:1993jj,Matsubara:1994nq,Bali:1998gz}.
The Abelian projected theory (AP-SU(2) gluodynamics) is closely
related to the dual superconducting scenario of the QCD
vacuum.
Abelian dominance of the string tension has left
no doubt that the vacuum has the property of dual superconductivity.
The dual Meissner effect, {\it i.e.} squeezing of the color-electric
flux into a string by the normal vacuum, can be studied both in
an effective dual Ginzburg-Landau (DGL) theory {\it and}
AP-SU(2) gluodynamics. Thus, one hopes to learn from real
gluodynamics about the vacuum properties
to be encoded in the DGL theory.

In papers~\cite{Bali:1998gz}
the structure of confining string was studied in details.
In this talk we report new results on the SU(2) flux-tube profile.
Our study extends previous work in various directions:
({\it i}) we reconsider the dependence on the quality of MAG fixing;
({\it ii}) we control the effect of smearing of the Wilson loops;
({\it iii}) we check the correct scaling behavior of the
flux-tube profile; and
({\it iv}) we investigate the detailed form
of the profile of the flux tube at finite distance between quark and
antiquark.
While ({\it i}) and ({\it ii}) mainly are an adaptation to present standards,
({\it iii}) was always badly missed.
Finally, ({\it iv}) makes possible a
detailed comparison of the finite-length flux tubes with the DGL theory.

The profile is described by correlation functions
\bea
\langle {\cal O}(s) \rangle_{W_A}
= \frac{\langle   W_{A} {\cal O}(s) \rangle_{0} }
{\langle  W_{A} \rangle_{0} },
\eea
where the source is the Abelian Wilson loop $W_{A}$ constructed out of
Abelian link variables. The local field operators ${\cal O}(s)$ of immediate
interest are the Abelian field strength $\bar{\theta}_{\mu\nu}(s)$ and the
monopole current $2\pi k_{\mu}(s)$ ($k_{\mu} \in Z\!\!\!Z$).
All are available after Abelian projection in MAG.
Correlation functions in terms of the photon part $W_{\mathit{Ph}}$ and the
monopole part $W_{\mathit{Mo}}$ instead of $W_{A}$ are
proposed~\cite{draft2},
which allow to discover the Coulomb and and
the solenoidal part of the field strength.

By using the Wilson gauge action,
SU(2) gauge field ensembles on a $32^4$ lattice have been generated at
$\beta=$ 2.3, 2.4, and 2.5115, each consisting of 100 configurations,
separated by 500 Monte Carlo sweeps
(after 2500 thermalization sweeps).
Corresponding gauge fixed ensembles have been considered and that
corresponding to the maximal value of MAG functional has been stored.
All calculations have been done at the Vector-Parallel Supercomputer
NEC SX-5 of RCNP, Osaka University.

We have found that the smearing parameters
$\alpha=2.0$, $N_{s}=8$ for the {\it spatial} Abelian links,
correspond to the optimal ground state overlap with $W_A$.
For calibration, the lattice spacing $a(\beta)$, has been found
from the relation $\sqrt{\sigma_{\mathit{phys}}} = \sqrt{\sigma_L}/a \equiv
440$ MeV.
Here $\sigma_L$ ($\sigma_{\mathit{phys}}$) is 
the lattice (physical)
string tension~\footnote{The lattice string tensions $\sigma_L$
are found to be 0.1373(16) at $\beta=2.3$,
0.0712(5) at $\beta=2.4$, and  0.0323(4) at $\beta=2.5115$,
which gives the lattice spacing
$a=0.1662(10)$ fm, $a=0.1197(4)$ fm,
and $a=0.0806(5)$ fm, respectively.}.
$\sigma_L$ has been extracted from
non-Abelian Wilson $W_{\mathit{NA}}$ loops using optimized
non-Abelian smearing.
The local energy is fitted in the form
$V(R) = C - A/R + \sigma_L R$.

\begin{figure}[!t]
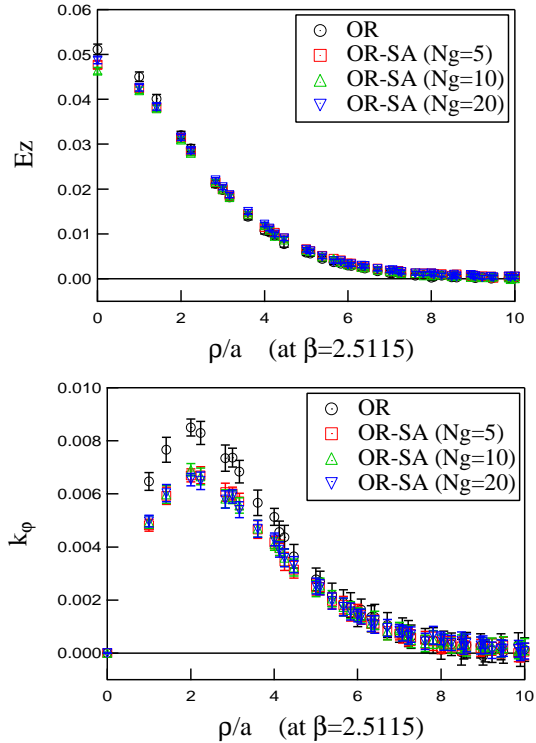

\includegraphics[height=5.cm]{E_ngdep.EPSF}
\includegraphics[height=5.cm]{k_ngdep.EPSF}
\vspace*{-1.5cm}
\caption{The electric field (upper) and monopole current (lower)
profiles for gauge fixing according to OR and OR-SA, respectively,
and the dependence on the number of gauge copies in the OR-SA case.
}\label{fig:ng_dep}
\vspace*{-0.5cm}
\end{figure}

In order to demonstrate the importance of the good gauge fixing,
we show in Fig.~\ref{fig:ng_dep} the profiles of electric field and monopole
current for $\beta=2.5115$ observed in the midplane between
$Q$ and $\bar{Q}$ at the distance $r=10a=0.81$
fm after MAG fixing.
We compare the overrelaxed steepest descent (OR) algorithm
with the OR-simulated annealing (OR-SA) algorithm. In the latter case,
we also study how the profile changes with the number of gauge copies.
For the electric profile the effect of insufficient
gauge fixing (OR) is limited
to an overestimation by less than 10 \%. The monopole current, however, is
strongly reduced by applying OR-SA. It is difficult to find a systematic
effect of $N_g$ ($N_g=$ 5, 10, 20) on the electric and monopole profiles.
It is absolutely necessary to choose the OR-SA algorithm but
a moderate number $N_g=5$ of Gribov copies seems to be sufficient.

\begin{figure}[!t]
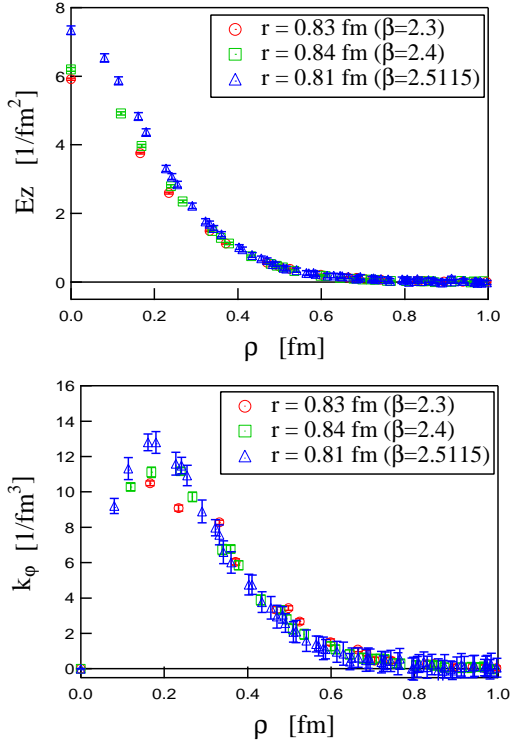

\includegraphics[height=5.cm]{E_scale.EPSF}
\includegraphics[height=5.cm]{k_scale.EPSF}
\vspace*{-1cm}
\caption{$\beta$ dependence of the profiles.}
\label{fig:physscale}
\vspace*{-0.5cm}
\end{figure}

In Fig.~\ref{fig:physscale} we plot the flux-tube profile in physical units
for three values of $\beta$. For each value of beta we choose
the (integer) lattice flux-tube length $R$ corresponding to
the (approximately) same $Q\bar{Q}$ distance $r \approx 0.8$ fm.  Both
profiles scale properly at $\rho \gtrsim 0.3$ fm.

\begin{figure}[!t]
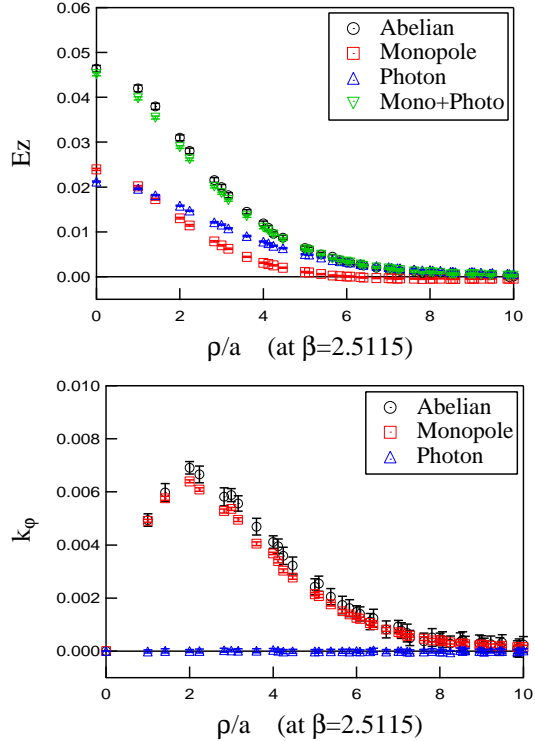

\includegraphics[height=5.cm]{E_structure.EPSF}
\includegraphics[height=5.cm]{k_structure.EPSF}
\vspace*{-1.3cm}
\caption{Decomposition of the flux-tube profiles
into photon- and monopole-induced parts.}
\label{fig:structure}
\vspace*{-0.3cm}
\end{figure}

In Fig.~\ref{fig:structure} we show
the profiles for $\beta=2.5115$ and a lattice inter-quark distance
$r=10a$, obtained from correlation functions between the local operators
${\cal O}(s)$ and the photon and the monopole parts of the Abelian Wilson
loops, respectively.  To accomplish this, the link field $\theta_l$ has been
first split into regular-photon and singular-monopole  parts.  For the
regular part of the links entering $W_{\mathit{Ph}}$, an Abelian smearing
similar to the original Abelian Wilson loop smearing has been found to be
optimal, giving an early plateau in $T$ for all $R$ of $\log
W_{\mathit{Ph}}(R,T)/W_{\mathit{Ph}}(R,T+1)$. Note that the
monopole part of the links has not been smeared.

We find that $W_{\mathit{Ph}}$ as source induces exclusively the
Coulombic electric field while the monopole part $W_{\mathit{Mo}}$
creates the solenoidal electric field.
Both contribute to the full
electric field produced by the full Abelian Wilson loop $W_{A}$.
The monopole current is not correlated with $W_{\mathit{Ph}}$.
Only the monopole part $W_{\mathit{Mo}}$ is responsible for
the (directed) circulating monopole current
induced in the vacuum.

We have compared both the electric field
and the monopole current observed in AP-SU(2) gluodynamics for a
$Q\bar{Q}$ system at finite separation with the classical
flux-tube solution obtained within the DGL theory.
Here, the aim is to determine the mass of the dual gauge boson
$m_{B}$ and the dual Higgs monopole field $m_{\chi}$,
and also the dual gauge coupling $\beta_{g}=1/g^{2}$.
We have formulated the DGL theory on the {\it dual lattice}
and have solved {\it the finite length flux-tube system}
numerically~\cite{Koma:2000hw}.
Preliminary fits of profiles at $\beta=2.5115$ for
$R$ = 4, 6 and 8 give the dual gauge boson mass around
1.0 GeV, which seems to be consistent with the value of $\rho$
corresponding to the maximum of the monopole current ($\rho \approx$
0.2 fm), see Fig.~\ref{fig:physscale}. The monopole mass was not
fixed.  The dual gauge coupling showed running behavior, depending on the
$Q\bar{Q}$ distance.  As increasing distance,  $\beta_{g}=1/g^{2}$ grows
gradually, which seems to follow the anti-screening behavior of the gauge
coupling $e$ in accordance with the Dirac quantization condition $eg=4\pi$.

\par
The authors are grateful to V.~Bornyakov, H.~Ichie, and G.~Bali
for useful discussions.
Y.~K. is partially supported  by
the Ministry of Education, Science, Sports and Culture,
Japan (Monbu-Kagaku-sho), Grant-in-Aid for Encouragement of Young
Scientists (B), 14740161, 2002.
E.-M.~I. gratefully acknowledges the support by the
Monbu-Kagaku-sho
and the hospitality at RCNP extended to him by H.~Toki.
M.~I.~P. is partially supported
by grants RFBR 02-02-17308, RFBR 01-02-117456, RFBR 00-15-96-786,
INTAS-00-00111, and CRDF award RPI-2364-MO-02.


\end{document}